# New exact spatially localized solutions of the (3 + 1) -dimensional nonlinear non-dissipative quasi-geostrophic potential vorticity equation for an exponential atmosphere

A.G. Kudryavtsev, N.N. Myagkov[1]

*Institute of Applied Mechanics, Russian Academy of Sciences, Moscow 125040, Russia*

**Abstract**

New exact spatially localized stationary solutions against the background of a zonal flow are found for the (3+1)-dimensional nonlinear non-dissipative quasi-geostrophic potential vorticity equation, which describes Rossby waves and vortices in an exponential atmosphere. In total, three solutions are presented. The nonlinear boundary conditions with a flat bottom and a rigid lid generate an infinite discrete set of baroclinic modes for each solution. The solutions show the possibility of existence of baroclinic dipoles in the exponential atmosphere, similar to baroclinic dipoles in the ocean. It is shown that: a) a pair of vortices in the baroclinic dipole appears and disappears when the velocity of stationary motion changes; b) the baroclinic dipoles show the ability to transfer warm or cold air depending on the polarity of the vortices in the dipole.

**Introduction**

The nonlinear non-dissipative quasi-geostrophic potential vorticity (QGPV) equation, which is also known as the Charney–Obukhov equation, is widely used to model the mesoscale circulations in the Earth's atmosphere and ocean [1, 2]. The QGPV equation, along with the Korteweg–de Vries, Boussinesq, nonlinear Schrödinger equations, etc., is a basic nonlinear equation, which is occurred not only in problems of geophysics, but also in problems of plasma physics (where it is known as the Hasegawa–Mima equation) [2, 3], mechanics of biological fluids [4], and in problems of astrophysics (e.g., [5,6]). In the 20th century, an efficient inverse scattering transform (IST) method [7] was developed to solve the above and other basic nonlinear equations, except for the QGPV equation. A necessary property of an equation integrable by the IST method is the presence of an infinite number of conservation laws. It was already noted in book [2] that the QGPV equation is Hamiltonian and admits the infinite number of conservation laws. However, the integrability of the QGPV equation was proven quite recently in work [8] where the Lax representation for this equation was found. However, the found Lax representation, valid only for a homogeneous atmosphere or ocean, has not yet led to the appearance of publications with new exact solutions using the IST method. Therefore, the problem of finding exact solutions to the HO equation using other methods remains relevant and of interest [9-13].

---
[1] Author to whom correspondence should be addressed: n.myagkov@iam.ras.ru

The currently known exact solutions of the QGPV equation for the atmosphere (see [2], [12-13]) are applicable only for the special case when a vertical perturbation size is much less than an atmosphere's scale height $\overline{H} = (-\partial \rho_s / \rho_s \partial z_*)^{-1}$, where $\rho_s(z_*)$ is the density of the undisturbed atmosphere and the asterisk indicates the dimensional coordinate. The new solutions of the QGPV equation presented in this paper are free from this drawback: they take into account the atmosphere inhomogeneity in the exponential approximation.

## 2. The solutions in the form of localized spherically symmetrical vortices against the background of a zonal flow.

The (3+1)-dimensional nonlinear QGPV equation in the *β*-plane approximation written in dimensionless variables has the form [1]

$$\frac{\partial}{\partial t}(\Delta \psi) + J(\psi, \Delta \psi) + \beta \frac{\partial}{\partial x}\psi = 0 \qquad (1)$$

where $\psi$ is the dimensionless geostrophic stream function; $\beta$ is the dimensionless meridional (northern) gradient of Coriolis parameter; $\Delta = \frac{\partial^2}{\partial x^2} + \frac{\partial^2}{\partial y^2} + \frac{1}{\rho_s(z)}\frac{\partial}{\partial z}\left(\frac{\rho_s(z)}{S}\frac{\partial}{\partial z}\right)$, where $\rho_s(z)$ is the density of the undisturbed atmosphere; $J(a,b) = \frac{\partial a}{\partial x}\frac{\partial b}{\partial y} - \frac{\partial a}{\partial y}\frac{\partial b}{\partial x}$ is the two-dimensional Jacobian. We assume that the troposphere of the earth's atmosphere can be characterized by an exponential decrease in density with height, $\rho_s(z) \propto \exp(-2hz)$, where $2h = H/\overline{H}$, $H$ – tropospheric height (distance from the earth's surface to the tropopause). Everywhere below, we assume that $H = 12$ km and $\overline{H} = 7.5$ km. The latter is assumed as the average value of $\overline{H}$ for the troposphere for mid-latitudes [14]. We believe that the parameter of stratification S in (1) is a constant parameter, and S =O(1) [1]. It is easy to see that it can be set to $S = 1$ without loss of generality. As usual, the *x*-coordinate is east, the y-coordinate is north, and the z-coordinate is up.

We have taken the boundary conditions with a flat bottom and a rigid lid as

$$\frac{d_0}{dt}\frac{\partial \psi}{\partial z} = 0 \text{ at } z = 0 \text{ and } z = H \qquad (2)$$

where $0 \leq z \leq H$ and $\dfrac{d_0}{dt} = \dfrac{\partial}{\partial t} + V_x \dfrac{\partial}{\partial x} + V_y \dfrac{\partial}{\partial y}$, $V_x = -\dfrac{\partial}{\partial y}\psi$, $V_y = \dfrac{\partial}{\partial x}\psi$. Boundary conditions (2) mean that the vertical velocity is zero at $z=0$ and $z=H$. The upper boundary (2) $z = H$ in atmospheric models is commonly associated with the level of tropopause that corresponds to a sharp jump in the gradient of the vertical background temperature profile (see, for example, [6]).

We are looking for solutions to the Eq. (1) with the boundary conditions (2) for stationary Rossby waves and vortexes propagating along the zonal direction at a constant velocity $V$: $\psi(t,x,y,z) = \psi(x - Vt, y, z)$.

Solution 1.

The equation (1) with the boundary conditions (2) has a solution in the form

$$\psi(t,x,y,z) = \dfrac{Ce^{hz}\sin(KR)}{R} - (V + U(z))y, \qquad (3)$$

where $R = \sqrt{(x-Vt)^2 + y^2 + z^2}$ and $C$ is an arbitrary constant,

$$U(z) = (A\sin(Kz) + B\cos(Kz))e^{hz} + \dfrac{\beta}{K^2 + h^2} \qquad (4)$$

$$A = \dfrac{\beta h}{K(K^2 + h^2)} \qquad (5)$$

$$B = \dfrac{\beta(\cos(KH) - e^{-hH})}{(K^2 + h^2)(\cos(KH) - e^{hH})} \qquad (6)$$

The wave number $K$ is the root of the equation that follows from the boundary conditions (2). This equation is

$$\sin(KH) = \dfrac{h(\cos(KH) - e^{hH})}{K} \qquad (7)$$

The boundary condition (2) on this solution has the form

$$\dfrac{d_0}{dt}\dfrac{\partial \psi}{\partial z} = CR^{-5}(x-Vt)e^{hz}(R^2(-hU(z) + \dfrac{d}{dz}U(z))(\sin(KR) - RK\cos(KR)) - \\ - zU(z)((K^2R^2 - 3)\sin(KR) + 3RK\cos(KR))) \qquad (8)$$

where $U(z)$ is defined by (4).

Solution 2.

The equation (1) with the boundary conditions (2) has a solution in the form (3), where $R = \sqrt{(x-Vt)^2 + y^2 + (z-H)^2}$; the zonal background flow is taken in the form (4), in which the value of parameter $A$ (5) remains the same, and parameter $B$ has the form

$$B = -\dfrac{\beta}{(K^2 + h^2)} \qquad (9)$$

The wave number $K$ is the root of the equation

$$\sin(KH) = -\frac{h(\cos(KH) - e^{-hH})}{K} \qquad (10)$$

The boundary condition (2) on this solution has the form

$$\frac{d_0}{dt}\frac{\partial \psi}{\partial z} = CR^{-5}(x-Vt)e^{hz}(R^2(-hU(z)+\frac{d}{dz}U(z))(\sin(KR)-RK\cos(KR)) - \\ -(z-H)U(z)((K^2R^2-3)\sin(KR)+3RK\cos(KR))) \qquad (11)$$

where $U(z)$ is defined by (4).

Solution 3.

The equation (1) with the boundary conditions (2) has a solution in the form

$$\psi(t,x,y,z) = \frac{C_1 e^{hz}\sin(KR)}{R} + C_2 e^{hz} J_0(Kr) - (V+U(z))y, \qquad (12)$$

where $R=\sqrt{(x-Vt)^2+y^2+z^2}$ and $r=\sqrt{(x-Vt)^2+y^2}$, $C_1$ and $C_2$ are the arbitrary constants, the zonal background flow is taken in the form (4), in which the value of parameter $A$ and $B$ are taken from (5) and (6). The wave number $K$ is the root of the equation (7).

The boundary condition (2) on this solution has the form

$$\frac{d_0}{dt}\frac{\partial \psi}{\partial z} = (x-Vt)e^{hz}\left(\frac{C_1 F_1(R,z)}{R^5} + \frac{C_2 F_2(r,z)}{r}\right)$$

$$F_1 = R^2(-hU(z)+\frac{d}{dz}U(z))(\sin(KR)-RK\cos(KR)) - zU(z)((K^2R^2-3)\sin(KR)+3RK\cos(KR))$$

$$F_2 = R^2(-hU(z)+\frac{d}{dz}U(z))J_1(Kr)$$

It is evident that Solution 3 is a partial superposition[2] of spherically and cylindrically symmetric solutions. Let us recall that for the case of a homogeneous ocean or atmosphere, there are exact solutions of the QGPV equation in the form of the partial superposition of the spherically symmetric solutions with vertically shifted centers [10]. Unfortunately, in the case of an exponential atmosphere, we were unable to find such solutions.

## 3. Model of the baroclinic dipole

It is evident that, similar to the solutions of the QGPV equation obtained in the constant density approximation [9-11], the exact solutions (3), (4) found for exponential atmosphere has the form

$$\psi(x-Vt,y,z) = C\varphi(x-Vt,y,z) + \Psi(y,z) \qquad (12)$$

---

[2] Partial superposition of solutions is a superposition that occurs only for the first terms on the right-hand side of formula (12), while the background flow remains unchanged.

where $\Psi(y,z) = -(V + U(z))y$, $V + U(z)$ - the velocity of the steady zonal background flow which depends on the velocity $V$ and parameters $K$, $H$, $\beta$ and $h$ included in the solution, and $C$ – arbitrary constant. Note that the second term on the right side $\Psi(y,z)$ in (1) is itself a solution of the QGPV equation for the case $C = 0$, but the first term $C\varphi(x - Vt, y, z)$ is not.

As we have already noted in [10], the vortices are localized in $z$ - neighborhood of the plane $z = z_c$, which is defined as a solution to the equation

$$V + U(z_c) = 0, \quad 0 \leq z_c \leq H \tag{13}$$

From equations (4)-(6) and (9), which define $U$, it is clear that the equation (13) gives the dependence of $z_c(V)$ for various values of parameters $K$, $H$, $\beta$ and $h$.

The solutions of equations (7) and (10) for given values of $H$, $\beta$ and $h$ give an infinite discrete set of the wave number $K$. The values of $K$ for the first few modes are shown in Table 1 for the parameters $h = 0.8$ and $H = \beta = 1$ (Figures 1-3 are plotted for the same parameter values $h = 0.8$ and $H = \beta = 1$). Equations (7) and (10) are invariant with respect to replacing $K \to -K$, therefore for every positive value of $K$ there is negative value $-K$. At the same time, as can be seen from (4)-(6) and (9), the zonal flow velocity is an even function of $K$, therefore the dependence of $z_c(V)$ extracted from (13) does not depend on the sign of $K$.

Table 1. The first eight values of the wave number $K$

| Mode # | Equation (7), $K$ | Equation (10), $K$ |
|---|---|---|
| 1 | 3.8215 | 2.7298 |
| 2 | 6.1206 | 6.2125 |
| 3 | 9.6913 | 9.3005 |
| 4 | 12.488 | 12.531 |
| 5 | 15.871 | 15.634 |
| 6 | 18.797 | 18.826 |
| 7 | 22.108 | 21.938 |
| 8 | 25.094 | 25.115 |

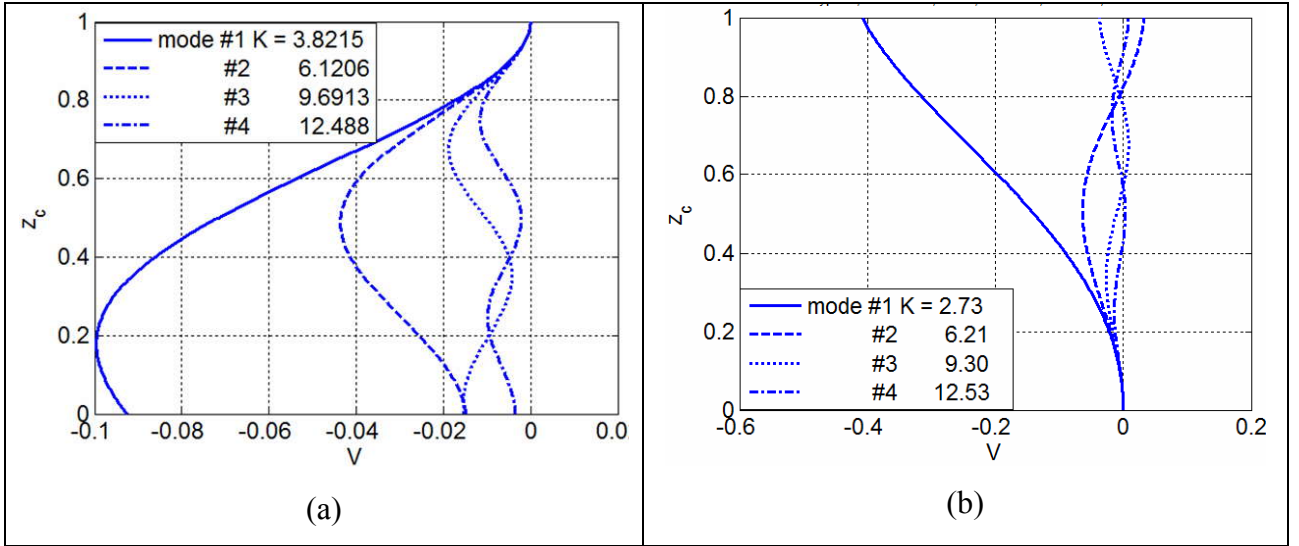

Fig. 1. The first four baroclinic modes for (a) Solution 1 and (b) Solution 2. The $K$ values for each mode are shown in the figures.

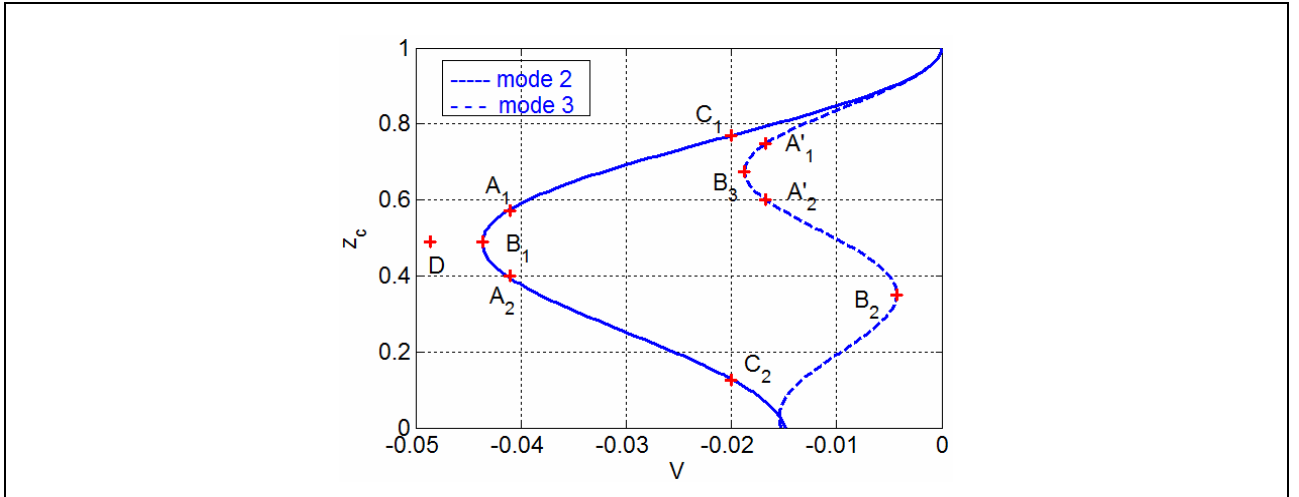

Fig. 2. The second and third baroclinic modes of Solution 1 with highlighted (+) points. The flows in the selected points $D(-0.0486, 0.49)$, $B_1(-0.0436, 0.49)$, $C_1(-0.02, 0.77)$ и and $C_2(-0.02, 0.126)$, are presented in Fig. 3(a,b,d); the dipole $A_1(-0.041, 0.57)$ - $A_2(-0.041, 0.4)$ is shown in Fig. 3(c); the dipole $A'_1(-1.67 \cdot 10^{-2}, 0.748)$ - $A'_2(-1.67 \cdot 10^{-2}, 0.6)$ is shown in Fig. 3(e).

Figures 1(a), (b) show the vertical structure of the first 4 baroclinic modes for Solutions 1 and 2. It is evident that for all cases (except mode 1 for Solution 2) in an exponential atmosphere there can be two or more vortices located at different heights and having the same velocity $V$. If two vortices are close enough to each other, they can form vertical (baroclinic) dipoles in which the vortices are connected by common streamlines. Let's look at this in more detail.

For the second and third modes of Solution 1, the dependences $z_c(V)$ are shown in Fig. 2 with selected points $D$, $B$, $A_1$, $A_2$, $C_1$, $C_2$ for the second mode and $B_1$, $B_2$, $A'_1$, $A'_2$ for the third mode; each of these points has coordinates ($V$, $z_c$) and corresponds to one vortex (excluding point D, which is located outside the curve $z_c(V)$), moving with velocity $V$ in the plane $z = z_c$. Points $B$ and $B_1$, $B_2$, located at the turning points, are critical points. A pair of vortices in a baroclinic dipole appears or disappears when the velocity parameter $V$ passes through these points. To the left of point $B$ in Fig. 2, i.e. at velocities $V < V_B$ (= - 0.0436), there are no baroclinic vortices. For example, at point D with $V_D$ = - 0.0486 there is a Rossby wave, which is shown in Fig. 3 (a). At $V = V_B$ there is one vortex, which is shown in Fig. 3(b), and at $V > V_B$ two vortices are born, which form the baroclinic dipole (for example, the vortices $A_1$ and $A_2$ in Fig. 2, which have the same velocity). With further increase of $V$ the vortices become independent of each other (vortices $C_1$ and $C_2$ in Fig. 2). The dipole $A_1$ - $A_2$ is shown in Fig. 3(c) and two isolated vortices $C_1$ and $C_2$ are shown in Fig. 3(d). We determine the existence of the dipole by the presence of common streamlines, which are highlighted in red and by dotted lines in Fig. 3. The value of the velocity parameter $V$, at which the dipole disintegrates into two independent vortices, is estimated as $V_{dis} \cong$ - 0.038. It is evident that the velocity interval $V_B < V < V_{dis}$, at which the dipole exists, is quite narrow. A similar picture occurs for the third mode shown in Fig. 2, where the dipole $A'_1$ - $A'_2$ is shown in Fig. 3(e).

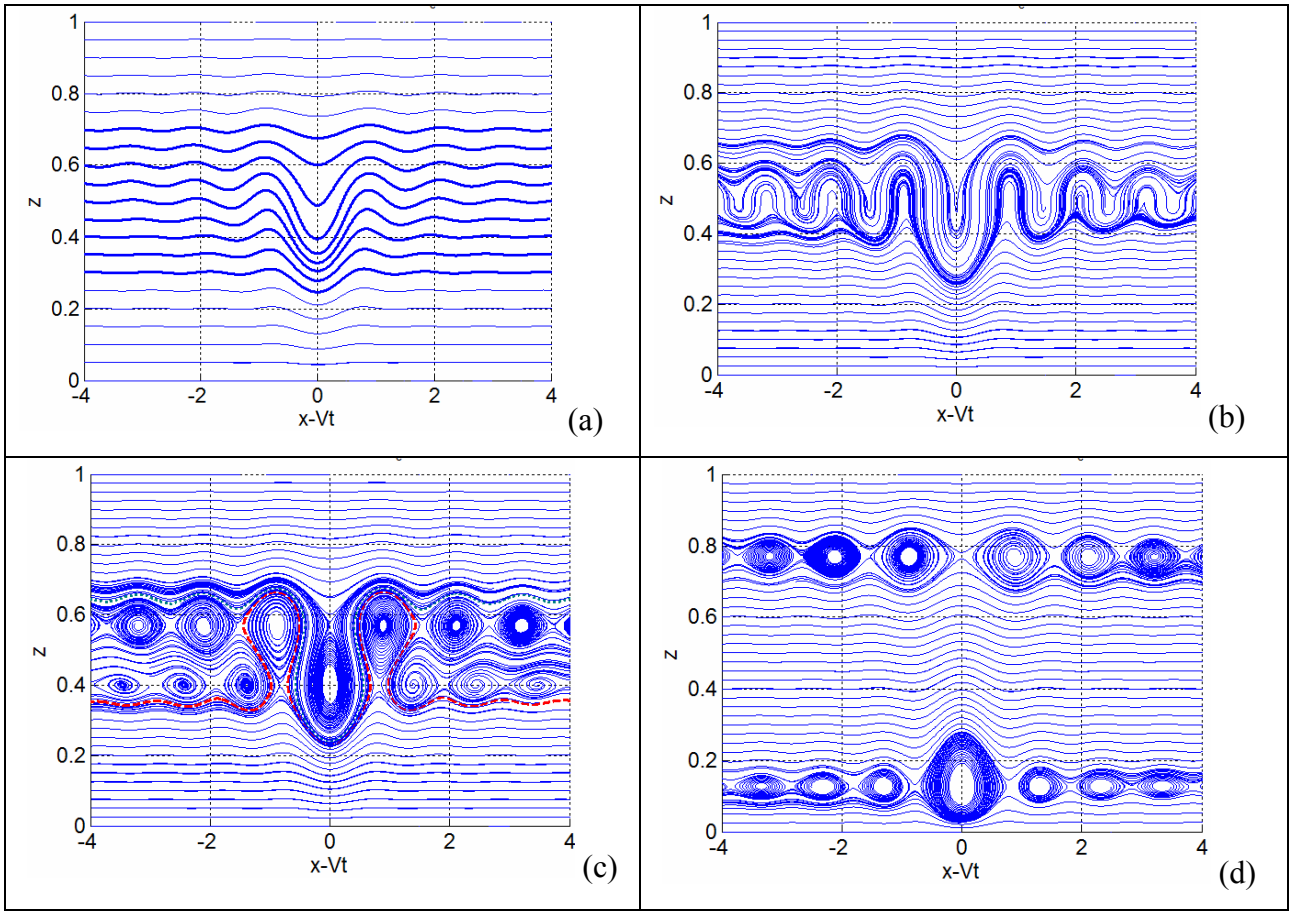

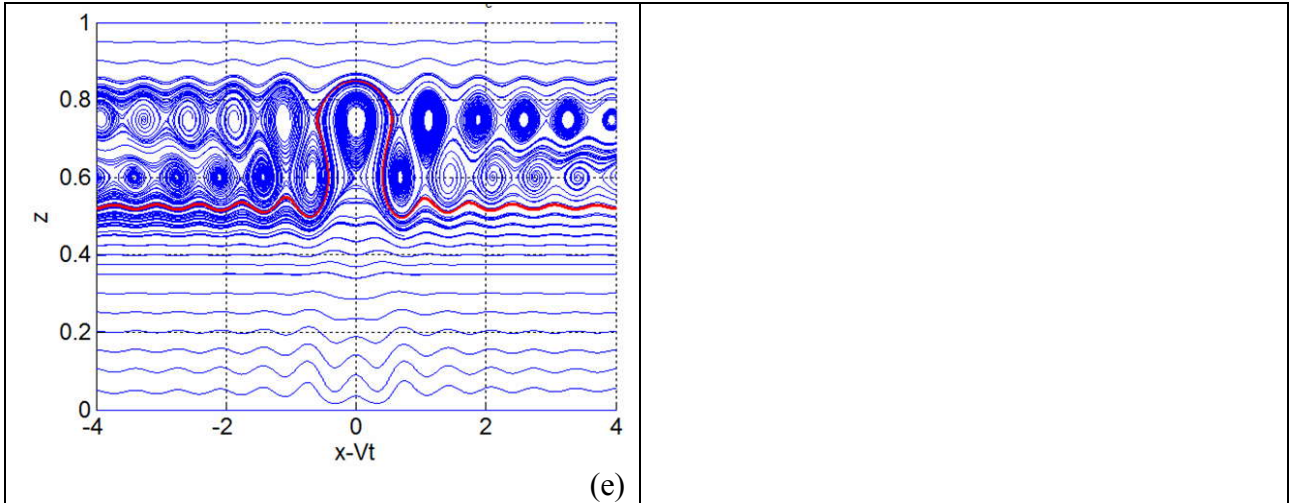

(e)

Fig. 3. Flow patterns of Solution 1 for second mode (a-d) and the third mode (e) in the *XZ* plane at section y = 0. The streamlines that are common to both vortices are shown in red and by dotted lines in figures (c) and (e). The parameter value $C = 10^{-3}$. Explanation in the text and in the caption to Fig.2.

The vortex at the critical point *B* (Fig. 3(b)) already has a large-volume "pocket" in the center, which is appeared during the formation of the baroclinic dipole and disappears at $V > V_{dis}$. Since in the real troposphere the temperature decreases with height down to the tropopause, the "pocket" can carry a large amount of cold air when moving in the troposphere. With reverse polarity of the vortices in the dipole, the "pocket" is directed upwards. This is realized, for example, for the dipole $A'_1$ - $A'_2$ of the third mode, which is shown in Fig. 3(e). In this case, the "pocket" can carry a large amount of warm air as the dipole moves. Similar baroclinic dipoles – hetons have previously been considered for oceanic eddies using approximate approaches and numerical modeling (see, for example, [15–17]).

**4. Conclusion**

Thus, for the (3+1)-dimensional nonlinear QGPV equation describing Rossby waves and vortices in an exponential atmosphere, new exact spatially localized (spherical) solutions are found against the background of a zonal flow, propagating along the zonal direction with a constant velocity *V*. Three solutions are presented. The nonlinear boundary conditions with a flat bottom and a rigid lid generate an infinite discrete set of the baroclinic modes for each solution. The first few modes of the vertical wave number *K* for each solution are discussed.

The solutions show that in the troposphere for a given value of the velocity parameter *V* there may be no baroclinic vortices, or there may be one, two or more vortices located at different heights and having the same velocity *V*. If a pair of such vortices is located close enough to each other, they can form vertical (baroclinic) dipoles. There are critical values of velocity $V$ that separate regions where there are no baroclinic vortex flows from regions where there are vortex flows in the form of baroclinic dipoles. In Fig. 2, the critical values of $V$ are at points $B$ and $B_1$, $B_2$.

Vortices in the baroclinic dipole form a "pocket" directed upward or downward depending on the polarity of the vortices in the dipole. Such the "pocket" is capable of capturing and transporting warm or cold air over long distances.